\documentclass{CSML}

\def\dOi{11(3:16)2015}
\lmcsheading%
{\dOi}
{1--16}
{}
{}
{Aug.~28, 2014}
{Sep.~21, 2015}
{}

\usepackage{amssymb}
\usepackage{hyperref}
\theoremstyle{plain}
\newtheorem{theoremy}{Theorem}

\newtheorem{exmp}[theoremy]{Example}

\theoremstyle{definition}

\newcommand{\seq}[1]{{\left\langle{#1}\right\rangle}}
\newcommand{\w}{\omega}
\newcommand{\dom}{\texttt{dom}}
\newcommand \inff{\texttt{inf}}
\newcommand \fin{\texttt{fin}}
\newcommand \outcome{\texttt{outcome}}
\newcommand{\converge}{\!\!\downarrow}
\newcommand{\diverge}{\!\!\uparrow}
\newcommand{\uhr}[1]{\! \upharpoonright_{#1}}

\begin{document}

\title[Partial functions and dominatio]{Partial functions and domination}
\thanks{{\lsuper a}C.~T.~Chong's research
was partially supported by NUS grant WBS C146-000-025-00.}
\thanks{{\lsuper c} F.~Stephan's
research was partially supported by NUS grants WBS R252-000-420-112,
WBS R146-000-181-112 and WBS R146-000-184-112 (MOE2013-T2-1-062).}

\author[C.~T.~Chong]{C.~T.~Chong\rsuper a}
\author[G.~Hoi]{Gordon Hoi\rsuper b}
\author[F.~Stephan]{Frank Stephan\rsuper c}
\author[D.~Turetsky]{Dan Turetsky\rsuper d}

\address{{\lsuper{a,b,c}}Department of Mathematics, National University of Singapore,
Singapore 119076, Republic of Singapore.}
\email{\{chongct, hoickg\}@math.nus.edu.sg}

\address{{\lsuper c}Department of Computer Science, National University of Singapore,
Singapore 117417, Republic of Singapore.}
\email{fstephan@comp.nus.edu.sg}

\address{{\lsuper d}Kurt G\"odel Research Center, W\"ahringer Stra\ss e 25,
1090 Wien, Austria.}
\email{dturets@gmail.com}


\ACMCCS{[{\bf Theory of computation}]: Models of computation---Computability---Recursive functions}

\begin{abstract}
\noindent
The current work introduces the notion of pdominant sets and studies
their recursion-theoretic properties. Here a
set $A$ is called pdominant iff there is a partial $A$-recursive
function $\psi$ such that for every partial recursive function $\phi$
and almost every $x$ in the domain of $\phi$ there is a $y$ in the
domain of $\psi$ with $y \leq x$ and $\psi(y) > \phi(x)$. While there
is a full $\Pi^0_1$-class of nonrecursive sets where no set is pdominant,
there is no $\Pi^0_1$-class containing only pdominant sets. No weakly
$2$-generic set is pdominant while there are pdominant $1$-generic sets
below $K$. The halves of Chaitin's $\Omega$ are pdominant.
No set which is low for Martin-L\"of random is pdominant.
There is a low r.e.\ set which is pdominant and a high r.e.\ set which
is not pdominant.
\end{abstract}

\maketitle

\section{Introduction}

\noindent
It is known since the beginning of recursion theory that there is
a close relation between the computational power of a set $A$
and the classes of recursive and partial recursive functions which
can be dominated by an $A$-recursive function \cite{Ma66a,Ma66b,Pos44}:
The halting problem $K$ is Turing reducible to $A$ iff every partial recursive
function is majorised by some $A$-recursive total function iff
some fixed $A$-recursive function dominates all partial recursive
functions. Furthermore, Martin showed that a set $A$ has high Turing
degree iff there is an $A$-recursive function $f$ dominating every
total recursive function. This generalises to r.e.\ Turing degrees
by the fact that an r.e.\ set $A$ is low$_2$ iff some $K$-recursive
function dominates all total $A$-recursive functions.

For r.e.\ sets, one has also the further result
that $A \leq_T B$ iff the convergence module of $A$
is majorised by some $B$-recursive function.
It is also known that for sets $A,B \leq_T K$, $A \leq_T B$
iff for every $A$-recursive function $f$ there is a $B$-recursive
function $g$ majorising $f$. This does, however, not generalise
to the larger class of the sets below $K'$: There is a nonrecurisve
set $A \leq_T K'$ for which every $A$-recursive function is majorised
by a recursive function; such a set $A$ is said to have a
hyperimmune-free degree \cite{MM68}.

There have also been various attempts to generalise domination to
relations between the partial recursive functions of
two relativised worlds. For example, Stephan and Yu \cite{SY14}
considered the notion where $A$ is strongly hyperimmune-free
relative to $B$; here $A$ is strongly hyperimmune-free relative
to $B$ iff for every partial $A$-recursive
function $f$ there is a partial $B$-recursive function $g$ such that
$dom(f) \subseteq dom(g)$ and $f(x) \leq g(x)$ for all $x \in dom(f)$.
Their results include the result that, when $B \not\geq_T K$ then
$A$ is strongly hyperimmune-fre relative to $B$ iff $A \leq_T B$.
So real power in terms of the degrees of being strongly hyperimmune-free
is only achieved by being above $K$. One of the reasons
for this phenomenon was that the domains have to be compatible in the
sense that whenever $f$ is defined so is $g$ and furthermore $g$ is
above $f$.

The notion investigated in the current work tries to overcome
this shortcoming by permitting $g$ to be above the value $f(x)$
not at $x$ itself but at places below $x$. In this initial work, the
emphasis is more on domination than majorisation; that is, on the
question of when there is a partial $B$-recursive function dominating
all partial $A$-recursive functions in this sense. Then $B$ must
be properly above $A$ and already the question when this can be
done with the lower oracle being recursive turned out to be an
interesting and challenging notion.

The notion itself stems from a principle in reverse mathematics,
more precisely, it is motivated by the bounded monotone enumeration
principle BME. Chong, Slaman and Yang \cite{CSY14} introduced BME
in order to analyse the rate of growth of a finite collection of partial
recursive functions in a nonstandard model of a fragment of Peano arithmetic.
The ability to control the growth rate in that model was key to
separating stable Ramsey's theorem for pairs from Ramsey's theorem
for pairs over the system RCA$_0$.

The formal definition of when a function pdominates another function
or a set of functions is the following.

\begin{defi}
If $\psi$ and $\phi$ are partial functions, $\psi$ pdominates $\phi$ iff
for almost all $x \in dom(\phi)$ there is a $y \in dom(\psi)$
with $y \leq x$ and $\psi(y) > \phi(x)$; $\psi$ pdominates a set of
partial functions iff it pdominates every partial function in it.
A set $A$ is pdominant iff some partial $A$-recursive function
pdominates the set of all partial recursive functions.
\end{defi}

\noindent
This property can indeed be formulated in a stricter way and therefore
the following equivalent property turned out to be fruitful for
various proofs.

\begin{prop} \label{pr:factork}
A set $A$ is pdominant iff there is a partial $A$-recursive
function $\psi$ such that for all constants $k$ and all
partial recursive functions $\phi$ and almost all $x \in dom(\phi)$
there is a $y \in dom(\psi)$ with $y \leq x/k$ and $\psi(y) > \phi(x)$.
\end{prop}

\proof
Given $\psi$ and a partial recursive function $\phi$ and a constant
$k$, consider the functions $\phi_\ell$ given by
$\phi_\ell(x) = \phi(kx+\ell)$. For each $\ell$ and almost all
$x$ in the domain of $\phi_\ell$ there is a $y \leq x$ in the
domain of $\psi$ with $\psi(y) > \phi_\ell(x)$. It follows that
for almost all $x$ in the domain of $\phi$ there is a $y \leq x/k$
in the domain of $\psi$ with $\psi(y) > \phi(x)$.\qed

\medskip
\noindent
Clearly the sets above the halting problem $K$ are pdominant.
One might ask whether there are natural examples of sets which
are pdominant but not above the halting problem. Indeed, taking
the half of Chaitin's $\Omega$ \cite{Cha75,Cha87,ZL70} provides such
an example. Martin-L\"of random sets \cite{ML66} are those sets $A$
for which there is not a uniformly r.e.\ sequence of open classes
$C_0,C_1,\ldots$ such that $\mu(C_n) < 2^{-n} \wedge A \in C_n$
for all $n$.

\begin{exmp} \label{ex:omega}
Let $\Omega$ be a left-r.e.\ Martin-L\"of random set and
$A$ be given by $A(n) = \Omega(2n)$. Then $A$ is low and
$A$ is pdominant.
\end{exmp}

\proof
The lowness of $A$ is well-known \cite[Corollary 3.4.11]{Nie09}.
Now it is shown that $A$ is pdominant. Let $\Omega_s$ be an approximation
of $\Omega$ from the left. On input $x = 2^n+\sum_{m<n} 2^m \cdot a_m$,
let $\psi^A(x)$ be the first stage $s$ such that
$\Omega_s(0)\Omega_s(1)\ldots \Omega_s(2n) =
  A(0)a_0A(1)a_1 \ldots A(n-1) a_{n-1}A(n)$.
The function $\psi^A$ is partial recursive and in the case that
$x = 2^n+\sum_{m<n} 2^m \cdot \Omega(2m+1)$, $\psi^A(x)$ is the time
to approximate the string $\Omega(0)\Omega(1)\ldots \Omega(2n)$ from the left.
Assume now by way of contradiction that there would be a partial
recursive function $\phi$ such that for infinitely many $x$ in the
domain of $\phi$ there is no $y \leq x$ with $\psi^A(y) > \phi(x)$.
Taking now $n$ such that $2^{n+1} \leq x < 2^{n+2}$ for such an
$x$, it would follow that $s = \phi(x)$ satisfies
$\Omega_s(m) = \Omega(m)$ for $m=0,1,\ldots,2n$. So one could
compute $\Omega$ up to $2n$ for infinitely many $n$ from an input
to $\phi$ which has $n+1$ binary bits. Hence, for infinitely many
$n$ it would hold that $C(\Omega(0)\Omega(1)\ldots \Omega(2n)) \leq n+c$
for some constant $c$ in contradiction to the fact that $\Omega$
is Martin-L\"of random.\qed

\medskip
\noindent
Note that in a similar spirit, one can show the following property:
If $A$ is a set which is Martin-L\"of random relative to $\Omega$,
then $\Omega$ is also Martin-L\"of random relative to $A$ and therefore
the convergence module $c_\Omega$ of $\Omega$ is a total function which
pdominates the set of all partial $A$-recursive functions. Thus, if
one relativises the notion, one can have two oracles, namely $A$ and
$\Omega$, such that a total $\Omega$-recursive function dominates all partial
$A$-recursive functions though $A \not\leq_T \Omega$.

For further properties of $\Omega$, see the usual textbooks on Kolmogorov
complexity and recursion theory \cite{DH10,LV08,Nie09,Odi89,Rog87,Soa87}.
In the following sections, it will be investigated which types of sets
can be pdominant and which not. The first results will deal with
$\Pi^0_1$ classes which are a convenient tool to construct sets of
various types; for the construction of pdominant sets, they turn out
to be of limited value as no $\Pi^0_1$ class consists entirely of
pdominant sets and some do not contain any pdominant set.

\section{$\Pi^0_1$-classes and pdominance}

\noindent
For the purposes of this paper, $\Pi^0_1$-classes are classes of
sets represented by their $\{0,1\}$-valued characteristic functions
and they are not empty. Furthermore, a $\Pi^0_1$-class is called
nonrecursive iff it contains only nonrecursive sets. $\Pi^0_1$-classes
correspond to the sets of infinite branches of recursive trees.

\begin{thm}
Every $\Pi^0_1$-class contains a set which is not pdominant.
\end{thm}

\proof
Let a $\Pi^0_1$-class be given as the infinite branches of a
recursive tree $T_0$. Now one defines inductively a sequence
of trees $T_1,T_2,\ldots$ such that the following holds for
all trees:
\begin{itemize}
\item $T_0 \supseteq T_1 \supseteq T_2 \supseteq \ldots$;
\item Each $e$ is linked to two numbers $a_e,b_e$ such that
      $T_e$ is computed by the function $\varphi_{a_e}$ and
      $b_e$ is the minimum possible size of the set
      $\{0,1,\ldots,2^{\langle a_e,e \rangle-2}\} \cap dom(\varphi_e^B)$
      when $B$ ranges over the infinite branches of $T_e$;
\item $T_{e+1}$ is a recursive tree having exactly those infinite branches
      of $T_e$ for which the intersection
      $\{0,1,\ldots,2^{\langle a_e,e\rangle -2}\}
      \linebreak[3] \cap dom(\varphi_e^B)$
      contains at most $b_e$ elements.
\end{itemize}
Note that the intersection of a descending sequence of binary trees
with infinite branches is not empty, as each tree defines a compact
subset of the set of possible infinite branches and the intersection
of a descending intersection of nonempty compact sets is not empty
in the Cantor Space. Let $A$ be an infinite branch which is on all
trees $T_e$.

Now consider the following function $\phi$: The algorithm for
$\phi(x)$ searches first for numbers $a,e,b$ with
$x = 2^{\langle a,e\rangle}+b$ and $b < 2^{\langle a,e \rangle}$.
If these numbers can be found, then the algorithm goes on with
determining for each set $B$ the first number $t$
found such that either $B(0)B(1)\ldots B(t)$ is not on the tree given
by $\varphi_a$ or there are $b$ distinct numbers
$u_1,u_2,\ldots,u_b \leq 2^{\langle a,e \rangle-2}$ satisfying
for each $y \in \{u_1,u_2,\ldots,u_b\}$ that
$\varphi^{B(0)B(1)\ldots B(t)}_{e,t}(y)$ is defined
and below $t$. In the case that these searches terminate for
all oracles $B$, for compactness reasons, the algorithm can find a
uniform proper upper bound $t'$ on all these $t$ and it then
outputs this~$t'$.

Assuming now by a way of contradiction that $\psi$ is the partial function
witnessing that $A$ is pdominant. Then there are infinitely many $e$ such that
$\varphi^A_e = \psi$. For each such $e$ and the corresponding $a_e,b_e$
it holds that (a) $\varphi^A_e$ is defined on $b_e$ places below
$2^{\langle a_e,e\rangle-2}$ for all infinite branches of $T_e$ and
(b) $\psi(2^{\langle a_e,e\rangle}+b_e)$ returns a value $t'$
which is larger than $\psi(y)$ for all $y$ in the domain of $\psi$
satisfying $y \leq 2^{\langle a_e,e\rangle-2}$; these $y$ contain every
$y \in dom(\psi)$ with $y \leq (2^{\langle a_e,e\rangle}+b_e)/8$. Hence
the function $\psi$ does not witness in sense of Proposition~\ref{pr:factork}
that $A$ is pdominant. As the choice of $\psi$ was arbitrary,
it follows that $A$ is not pdominant.\qed

\begin{thm}
There is a nonrecursive $\Pi^0_1$-class not containing any
pdominant set.
\end{thm}

\proof
The idea is to construct a partial recursive $\{0,1\}$-valued
function $\vartheta$ and let the $\Pi^0_1$-class be the class
of all total extensions. At each stage $s$, let $a_{m,s}$ be
the $m$-th non-element of the domain of $\vartheta_s$.
Now one updates $\vartheta$ according to that of the following
actions which has highest priority, that is, the lowest parameter
$m$ being involved.
\begin{itemize}
\item If $\varphi_{m,s}$ and $\vartheta_s$ are consistent, that is,
      not different in the common domain, and if $\varphi_{m,s}(a_{m,s})$
      is defined, then make $\vartheta_{s+1}(a_{m,s})$ be defined and
      different from $\varphi_{m,s}(a_{m,s})$.
\item If there is a string $\tau \in \{0,1\}^s$ consistent with
      $\vartheta_{m,s}$ and the value
      $$
         u = \langle m,a_{0,s},\tau(a_{0,s}),a_{1,s},\tau(a_{1,s}),
         \ldots,a_{m,s},\tau(a_{m,s}) \rangle
      $$
      does not satisfy that
      $\varphi^{\tau}_{e,|\tau|}(y) \downarrow \Rightarrow
      \varphi^{\tau}_{e,a_{m+1,s}}(y) \downarrow$ for all $e,y \leq 2^{u-2}$
      then let $\vartheta_{s+1}(z) = \tau(z)$ for
      $z = a_{m,s}+1,a_{m,s}+2,\ldots,s-1$.
\end{itemize}
Here it is assumed that at any pairing of finitely many coordinates
the formed pair is larger than every single coordinate.
Note that the requirements are set up in a way that it happens
only at finitely many stages that $\vartheta_{s+1}(a_{m,s})$ becomes
defined, hence for each $m$ the limit $a_m$ of the $a_{m,s}$
exists and is outside the domain of $\vartheta$.
By the first of the two requirement for $m$, every member of the
resulting $\Pi^0_1$-class will be different from $\varphi_m$
whenever the latter is total. So the resulting $\Pi^0_1$-class
is nonrecursive.

Now assume by way of contradiction that $A$ is in the resulting
$\Pi^0_1$-class and that $\psi = \varphi^A_e$ witnesses that $A$ is
pdominant. Then there is, for each $m > e$, a value $u$ of the form
      $$
         u = \langle m,a_0,A(a_0),a_1,A(a_1),
         \ldots,a_m,A(a_m)\rangle.
      $$
Let $b$ be the number of places $y$ such that $\varphi^A_e(y)$
is defined below $2^{u-2}$. Now one can define a partial recursive
function $\phi$ such that whenever $x = 2^u+b$ for $u,b$ as above
then $\psi(x)$ decodes $m,a_0,a_1,\ldots,a_m,A(a_0),A(a_1),\ldots,A(a_m)$
from $u$ and waits until an $s$ is found such that $a_{k,s} = a_k$ for
$k \leq m$ and there are $b$ values $y \leq 2^{u-2}$ for which
$\varphi^A_{e,a_{m+1,s}}(y)$ is defined and is below $s$.
Then $\phi(x)$ takes the value $s+1$.

One can easily verify that whenever the parameters $e,b$ are taken
as above, all the involved computations halt and the corresponding $s$
is found and the value $s+1$ returned is strictly larger than $\psi(y)$ for
all $y \leq x/8$ which are in the domain of $\psi$. Hence, by
Proposition~\ref{pr:factork}, $A$ cannot
be pdominant in contradiction to the assumption. In other words, no
member $A$ of the constructed $\Pi^0_1$-class is pdominant.\qed

\begin{rem}
One can construct a $\Pi^0_1$-class with one nonrecursive branch of
the same Turing degree as $K$ and some recursive branches. Hence there
is a nontrivial $\Pi^0_1$-class where all nonrecursive members are
pdominant.
\end{rem}

\section{Pdominant sets and genericity}

\noindent
An extension function $F$ is a partial function which on input $\sigma$
is either undefined or maps $\sigma$ to a string extending $\sigma$.
A set $A$ meets an extension function $F$ iff there is an $n$ such
that either $F(A(0)A(1)\ldots A(n) \sigma)$ is undefined for all $\sigma$
or $F(A(0)A(1)\ldots A(n)) = A(0)A(1)\ldots A(m)$ for some $m \geq n$.

A set $A$ is called {\em $1$-generic} iff it meets every partial recursive
extension function; a set $A$ is called {\em weakly $2$-generic} iff
it meets every total $K$-recursive extension function.

Note that every weakly $2$-generic set meets each total $K$-recursive
extension function infinitely often in the sense that for every such $F$
there are infinitely many $n$ with $F(A(0)A(1)\ldots A(n))$ being a
prefix of $A$.

{\sloppy The $1$-generic sets are used implicitly in Friedberg's Jump Inversion
theorem, see \cite{Fr57}.  Jockusch \cite{Jo77,Jo81} formally introduced
$1$-generic sets and its variants.}

\begin{thm}
No weakly $2$-generic set is pdominant.
\end{thm}

\proof
Let $A$ be a weakly $2$-generic set. Assume by way of contradiction
that $A$ is pdominant and that the partial $A$-recursive function
$\psi^A$ with oracle $A$ witnesses this fact.

Define a partial recursive extension function
$\gamma$ such that $\gamma(\sigma,u)$ is the length-lexicogra\-phically
first extension $\tau$ of $\sigma$ such that there is an
$x \leq u$ for which $\psi^{\tau}_{|\tau|}(x)$ is defined
while $\psi^{\sigma}_{|\sigma|}(x)$ is undefined.
$\Gamma(\sigma,u)$ is obtained by iterating $\gamma$
as long as possible: $\Gamma(\sigma,u) = \sigma_n$ for the
largest $n$ where $\sigma_n$ is defined where inductively
$\sigma_0 = \sigma$ and $\sigma_{m+1} = \gamma(\sigma_m,e,u)$
whenever that is defined. Note that $\gamma$ is partial recursive
and that $\sigma_{u+2}$ cannot exist as for each $m$ where
$\sigma_{m+1}$ is defined there is at least one $x_m \leq u$ with
$\psi^{\sigma_k}_{|\tau|}(x_m)$ being defined iff $\sigma_k$
exists and $k \geq m+1$. Hence $\Gamma$ is a total $K$-recursive
function.

Now one considers the extension function
which maps $\sigma$ to $\Gamma(\sigma,2^{no(\sigma)})$
where $no$ is a recursive bijection from strings to natural
numbers.

There are infinitely many prefixes $\sigma_0$ of $A$ such that
$\Gamma(\sigma_0,2^{no(\sigma_0)-2})$ is a prefix of $A$ as well.
For any such prefix $\sigma_0$, let $c = no(\sigma_0)-2$ and let
$n$ be the maximal number such that $\sigma_n$ is defined where
$\sigma_{m+1} = \gamma(\sigma_m,2^{c-2})$ whenever $\sigma_m$
is defined and $\gamma(\sigma_m,2^{c-1})$ is defined.

Let $\phi(2^c+n)$ be the partial recursive function
which computes $c,n,\sigma_0$ from $2^c+n$, computes $\sigma_n$ from
$\sigma_0$ through $n$ times iterating $\gamma$ with the corresponding
inputs and then outputs
$$
   1+\max\{\psi^{\sigma_n}_{|\sigma_n|}(x):
     x \in \mbox{dom}(\psi^{\sigma_n}_{|\sigma_n|}) \wedge x \leq 2^{c-2}\}.
$$
One can now easily see that $\phi(2^c+n) > \psi(y)$ for all $y \leq 2^{c-2}$
in the domain of $\psi$ in the case that
$\sigma_0,\Gamma(\sigma_0,2^{no(\sigma_0)-2})$
are both prefixes of $A$ and $n$ is the
number of extensions such that $\sigma_n = \Gamma(\sigma_0)$.
So, for the constant $k = 8$ and the infinitely many $x$ of the form
$2^c+n$ in the domain of $\phi$
with $n,c$ derived from the corresponding $\sigma_0$ there
is no $y \leq x/8$ in the domain of $\psi$ with $\psi(y) > \phi(x)$.
Hence $A$ is not pdominant.\qed

\begin{thm} \label{th:pdomonegeneric}
There is a pdominant $1$-generic set.
\end{thm}

\proof
The idea is to construct the $1$-generic set $A$ as a sequence
$\sigma_0 \tau_0 \sigma_1 \tau_1 \ldots$ such that
the $\sigma_n$ is chosen such that the $n$-th $1$-genericity
requirement is satisfied and $\tau_n$ contains information
about the halting problem $K$ which is sufficient to permit
partial domination.
More precisely, $\sigma_n$ is the first string found such that
$\sigma_0 \tau_0 \sigma_1 \tau_1 \ldots \sigma_{n-1}\tau_{n-1} \sigma_n$
is enumerated into the $n$-th r.e.\ set of strings; if no such string
is found then one takes $\sigma_n$ to be the empty string.
These enumeration procedures are fixed from now on, so that the
coding and decoding always refers to the same way of enumerating the $n$-th
r.e.\ set. The string $\tau_n$ encodes
whether $\varphi_e(x)$ halts for all $e,x < 2^{n+3}$:
So $\tau_n$ has length $4^{n+3}$ and for all $e,x < 2^{n+3}$, 
if $\varphi_e(x)$ halts then $\tau_n(2^{n+3}e+x) = 1$
else $\tau_n(2^{n+3}e+x) = 0$.

Now one can construct the following function $\psi^A$ using the
algorithm below which is undefined whenever some of the computations on the
way are not terminating:
\begin{itemize}
\item On input $y$, determine $n,a_0,\ldots,a_n$ such that
      $y = 2^{n+1}+\sum_{m \leq n} 2^m a_m$.
\item For $m=0,1,\ldots,n$: if $a_m = 1$ then search for the first $\sigma_m$
      found such that the string
      $\sigma_0 \tau_0 \sigma_1 \tau_1 \ldots \sigma_{m-1}\tau_{m-1} \sigma_m$
      is enumerated into the $n$-th r.e.\ set
      else let $\sigma_m$ be the empty string; let $\tau_m$ be that string
      of length $4^{m+5}$ which makes
      $\sigma_0 \tau_0 \sigma_1 \tau_1 \ldots \sigma_m \tau_m$
      to be a prefix of $A$.
      If either the search does not terminate or the prefix-condition
      cannot be satisfied then $\psi^A(y)$ is undefined.
\item Having $\tau_n$, one simulates all computations $\varphi_e(x)$
      with $e,x < 2^{n+3}$ and $\tau_n(2^{n+3}e+x)=1$
      until they terminate with a value $z$.
      If all these simulations terminate then $\psi^A(y)$ is $z+1$
      for the largest $z$ returned in these simulations else
      $\psi^A(y)$ is undefined.
\end{itemize}
If now $\varphi_e(x)$ is defined, $n>e$ and $2^{n+2} \leq x < 2^{n+3}$ then
one can choose $a_0,a_1,\ldots,a_n$ such that $a_m = 1$ whenever the
search for $\sigma_m$ finds this string and $a_m = 0$ otherwise.
Now considering $y = 2^{n+1}+\sum_{m \leq n} 2^m \cdot a_m$,
$\psi^A(y)$ is defined, $\psi^A(y) > \varphi_e(x)$ and
$y < x$. Hence $A$ is pdominant.\qed

\medskip
\noindent
Note that the $1$-generic set constructed is actually below $K$.
Theorem~\ref{th:lowforml} below shows that there are also
$1$-generic sets below $K$ which are not pdominant; this holds
as there are $1$-generic sets which are low for Martin-L\"of
random.

\section{Pdominance and lowness}

\noindent
As mentioned in the introduction, a pdominant set $A$ can be low,
that is, satisfy $A' \leq_T K$.

Furthermore, it can even be superlow, as one can construct
the sequence of the $\sigma_n$ and $\tau_n$ relative to
Theorem~\ref{th:pdomonegeneric} such that the $\sigma_n$ is not
taken to satisfy a $1$-genericity requirement but to fix the
$n$-th bit in the jump of $A$. Indeed, one can for each $n$
ask how often approximations to the sequence of the $\sigma_0 \tau_0
\sigma_1 \tau_1 \ldots \sigma_n \tau_n$ changes where each change
is either caused by some bit in some $\tau_m$ going from $0$ to $1$
or by finding a $\sigma_m$; the latter can be undone later if some
$\sigma_k$ is found with $k < m$ or some bit is enumerated into
a $\tau_k$ with $k < m$. Nevertheless, one can compute from $n$
an upper bound on the number of changes and a truth-table query
to $K$ permits then to get the amount of actual changes.
Hence $A'$ is truth-table reducible to $K$.

So the next step in making pdominant sets to be even lower than
superlow would be to make them low for Martin-L\"of random.
But that goal can no longer be achieved as the following
result shows.

\begin{thm} \label{th:lowforml}
Let $A$ be a pdominant set. Then $A$ is not low for Martin-L\"of randomness.
\end{thm}

\proof
Recall that $H$ denotes the prefix-free Kolmogorov complexity \cite{Ca02}.
First, it is well-known that one can choose the universal machine $U$ such that
there is a recursive approximation $H_t$ of $H$ from above satisfying for
infinitely many $x$ that $H(x) = H_0(x)$. In other words, the first
approximation $H_0$ is a Solovay function for $H$ \cite[Proof of
Proposition 2.5]{FSW06}.

Second, for each $x$ let $s(x,c)$ denote the number of strings $y$ which
are of length $x$ such that $H(y) < x+H_0(x)-c$. Fix the value c such that
there are infinitely many $x$ with $H(x) = H_0(x)$ and $s(x,c) < 2^x$: This
value exists, since there is a constant $c'$ with
$$
   \sum_{y \in \{0,1\}^x} 2^{-H(y)} > 2^{-c'-H(x)}
   \mbox{ and } \forall y \in \{0,1\}^x\,[H(y) < H(x)+c'].
$$
The assumption $H(y) < x+H_0(x)-c'$ for all $y$ of length $x$
implies that
$$
   \sum_{y \in \{0,1\}^x} 2^{-H(y)} \leq
   \sum_{y \in \{0,1\}^x} 2^{-x-H_0(x)-c'} \leq 2^{-c'-H_0(x)}
$$
and thus $H(x) < H_0(x)$. It follows that any $c>c'$ is a reasonable
choice to satisfy the condition, so let $c = c'+2$. For this $c$, there
is furthermore a minimal constant $d$ such that $s(x,c) \geq 2^{x-d}$ for
infinitely many $x$ satisfying $H(x) = H_0(x)$ and $s(x,c) < 2^x$:
As mentioned, all strings $y$ of length $x$ satisfy $H(y) \leq H(x)+c'$.
Thus there is a constant $c''$ such that for all $d'$, $2^{x-2d'}$ of
these strings $y$ satisfy $H(y) \leq H_0(x)+x+c''-d'$. From this fact,
one can conclude the existence of some $d$ with the desired properties
and then take the minimal one among the possible values. Let
$$
   B = \{x: H(x) = H_0(x) \wedge 2^{x-d} \leq s(x,c) < 2^{x-(d-1)}\}
$$
be the infinite set of these $x$.

Third, by Proposition~\ref{pr:factork},
$\psi^A$ is a partial $A$-recursive function such that for
all $\varphi_e$, all $k$ and almost all $a \in dom(\varphi_e)$ there is a
$b \in dom(\psi^A)$ with
$b \leq a \cdot 2^{-2k-1} \wedge \psi^A(b) > \varphi_e(a)$.

Fourth, now let $\varphi_e$ be the function which on input $a$ computes
first the integer $x$ with $2^{x-d} \leq a < 2^{x-(d-1)}$ and then searches
for the first $t$ such that there are $a$ distinct strings $y \in \{0,1\}^x$
with $H_t(y) < x+H_0(x)-c$. Note that for all $x \in B$ the value
$\varphi_e(s(x,c))$ is defined and returns the time $t$ which is needed
until all strings $y \in \{0,1\}^x$ with $H(y) < x+H_0(x)-c$ indeed satisfy
$H_t(y) < x+H_0(x)-c$. For each number $k$ and almost all $x \in B$
there is a $b \leq a \cdot 2^{-2k-1}$ with $\psi^A(b) > \varphi_e(s(x,c))$;
note that $b \leq a \cdot 2^{-2k-1}$ implies $b < 2^{x-2k-1}$.

Fifth, consider any $x$ and let $p$ be any program for the universal machine
on which $H$ is based with output $x$. Now one makes a prefix-free
machine $M^A$ which takes inputs of the form $p \cdot 0^k1 \cdot z$ where
$p$ is in the domain of the prefix-free universal machine and the length
of $z$ is $x-2k-1$ provided that the prefix-free universal machine outputs $x$
on input $p$. On inputs of these form, $M^A$ finds $p$ and determines $x$,
then $M^A$ determines the binary value $b$ of $z$ and then $M^A$ runs the
computation $t = \psi^A(b)$. If all these computations halt then $M^A$
outputs the lexicographic first string $y \in \{0,1\}^x$ with
$H_t(y) \geq x+|p|-c$ (provided that this exists).

Sixth, in the case that $x \in B$ and $p$ is a shortest program for $x$
in the universal machine and $\psi^A(b) > \varphi_e(s(x,c))$ for some
$b < 2^{x-2k-1}$ and $z$ coding $b$ in $x-2k-1$ binary bits,
$M^A(p \cdot 0^k1 \cdot z)$ returns a string $y$ with $H(y) \geq x+H(x)-c$.
Furthermore, the input for $M^A$ to generate $y$ is of length $|p|+x-k$
where $|p| = H(x)$.
It follows that there is for each $k$ a string $y$ such that the partial
$A$-recursive machine $M^A$ generates $y$ from a string which is $k-c$
bits shorter than its unrelativised Kolmogorov complexity.
Hence $A$ is not low for prefix-free Kolmogorov complexity and
thus $A$ is also not low for Martin-L\"of randomness.\qed

\medskip
\noindent
The last result showed that every set which is low for Martin-L\"of random
cannot be pdominant. As there are sets of nonrecursive r.e.\ degree,
sets of minimal Turing degree and $1$-generic sets which are all
low for Martin-L\"of random, one gets by this result alternative
proofs to the preceding ones that sets of those types might not
be pdominant.

One generalisation of sets which are low for
Martin-L\"of random are those which are low for $\Omega$. Therefore
the following question is natural.

\begin{oprob}
Is there a pdominant set which is low for $\Omega$?
\end{oprob}

\noindent
Furthermore, one might also like to know whether a further
lowness notion is compatible with pdominance.

\begin{oprob}
Is there a pdominant set of hyperimmune-free Turing degree?
\end{oprob}

\section{R.e.\ sets and pdominance}

\noindent
In this section it is shown that pdominant sets are orthogonal to the
low-high-hierarchy. That is, there is a low r.e.\ set which is pdominant
and a high r.e.\ set which is not pdominant.

\begin{thm} \label{th:pdomrelow}
There is a low r.e.\ set which is pdominant.
\end{thm}

\proof
The idea would be to construct in the limit a set $A$ such that
its characteristic function consists of a sequence
$\sigma_0 \tau_0 \sigma_1 \tau_1 \ldots$ where each $\sigma_n$
can become longer over time and each $\tau_n$ has the length
$2^{n+3}$. As in Theorem~\ref{th:pdomonegeneric}, $\tau_n(2^{n+3}e+x) = 1$
if $\varphi_e(x)$ halts and $\tau_n(2^{n+3}e+x) = 0$ if
$\varphi_e(x)$ does not halt for all $e,x < 2^{n+3}$. Furthermore,
the $\sigma_n$ can be made longer and takes then the current values of
$A_s$ at the corresponding positions, that is, increasing the length
of $\sigma_n$ does never change the values of $A_s$.
Whenever an interval $\sigma_n$ becomes longer than it is made to extend
up to $s$ and the strings $\tau_n,\sigma_{n+1},\ldots$ keep their length
but have all bits reset to $0$, as $A_s$ does not contain elements
beyond $s$. The algorithm is now the following:
\begin{itemize}
\item Identify the first string $\sigma_n$ or $\tau_n$ in the
      characteristic function which needs attention.
\item Here $\sigma_n$ occupying an interval from $i$ to $j$ with
      $j<s$ needs attention when $\varphi^{A_s}_n(n)$ converges
      but needs more than $j$ and less than $s$ steps.
      If $\sigma_n$ receives attention then $\sigma_n$ is adjusted
      to cover the interval from $i$ to $s$ and $A_{s+1} = A_s$,
      that is, $\sigma_n$ takes the value $A_s(i)A_s(i+1)\ldots A_s(s)$.
      Each of the strings $\tau_n,\sigma_{n+1},\tau_{n+1},\sigma_{n+2},
      \tau_{n+2},\ldots$ keeps it length but has its values set to
      $0$, so if $\tau_{n+1}$ was $b_1 b_2 \ldots b_k$ for some $k$
      then its new value is $0^k$.
\item Here $\tau_n$ occupying an interval from $i$ to $j$ with $j<s$
      needs attention when $\tau_n(2^{n+3}e+x)=0$ and $\varphi_e(x)$
      converges within $s$ computational steps.
      If $\tau_n$ receives attention then $i+2^{n+3}e+x$ is put into $A_{s+1}$
      and thus $\tau_n(2^{n+3}e+x)$ becomes $1$. Furthermore, if
      $\sigma_{n+1}$ does not cover $s$ then it is enlarged and takes
      the value $A_s(j+1)A_s(j+2)\ldots A_s(s)$ and
      each of the strings $\tau_{n+1},\sigma_{n+2}, \tau_{n+2},\sigma_{n+3},
      \tau_{n+3},\ldots$ keeps it length but has its values reset to
      $0$, so if $\tau_{n+1}$ was $b_1 b_2 \ldots b_k$ for some $k$
      then its new value is $0^k$.
\end{itemize}
Note that when $\sigma_0,\tau_0,\sigma_1,\tau_1,\ldots,\sigma_{n-1},\tau_{n-1}$
have found their final values then $\sigma_n$ will receive attention at most
once in order to change $A'(n)$ from $0$ to $1$; afterwards $\tau_n$ might
receive attention up to $4^{n+3}$ times as each of its bits can go from $0$
to $1$.

Note that $A'(n)$ can only change from $1$ to $0$ when some string of
priority higher than $\sigma_n$ receives attention or when the right end
of $\sigma_n$ is beyond $s$. This happens only finitely often and so $A$
is low.

Now define $\psi^A(y)$ as follows. Determine $n,a_0,a_1,\ldots,a_n$
such that $y = 2^{n+1}+\sum_{m\leq n}2^m \cdot a_m$ (if these values exist).
Now find the first $s$ such that, inductively, the following holds
for $m=0,1,\ldots,n$: $\sigma_m$ has received exactly $a_m$ times
attention after the last time that some interval before $\sigma_m$
has received attention; the string $\sigma_0 \tau_0 \sigma_1 \tau_1
\ldots \sigma_n \tau_n$ is a prefix of $A$. If such a stage $s$
is found, then simulate all computations $\varphi_e(x)$ with
$e,x < 2^{n+3}$ until they deliver an output $z$ and return the least
proper upper bound of these values $z$.

Now consider any $\varphi_e$ and any $n,x$ with $e < n$ and
$2^{n+2} \leq x < 2^{n+3}$. Furthermore, choose $a_0,a_1,\ldots,a_n$
such that $a_m = 1$ iff in the construction $\sigma_m$ has received
attention after all the higher priority strings $\sigma_0,\tau_0,
\sigma_1,\tau_1,\ldots,\sigma_{n-1},\tau_{n-1}$ have received attention
for the last time. Then, in the above definition,
$\psi^A(y)$ with $y = 2^{n+1}+\sum_{m\leq n}2^m \cdot a_m$
is defined and is a proper upper bound of $\varphi_e(x)$ whenever
the latter is defined. This witnesses that $A$ is pdominant.\qed

\begin{rem}
Actually one can show in Theorem~\ref{th:pdomrelow}
that $A$ is even superlow. This is done by
observing that a string $\sigma_n$ receives attention at most once
in any interval of time where no higher priority string receives
attention. Furthermore, $\tau_n$ receives attention at most
$4^{n+3}$ times during any interval of time where no higher priority
string receives attention. This permits to conclude that the strings
$\sigma_0,\tau_0,\sigma_1,\tau_1,\ldots,\sigma_n,\tau_n$ altogether
receive at most
$2^{n+1} \cdot 5^3 \cdot 5^4 \cdot 5^5 \cdot \ldots \cdot 5^{n+3}$
times attention. One can therefore find out with a tt-reduction to
$K$ how often these intervals receive attention and then simulate
the construction until that attention was granted for the last time
and then one knows whether $A'(n)$ holds or not. So $A' \leq_{wtt} K$
and indeed $A' \leq_{tt} K$ as wtt-reductions to $K$ can be turned
into tt-reductions. It follows that $A$ is superlow.
\end{rem}

\begin{thm}\label{T:5.3}
There is a high r.e.\ set which is not pdominant.
\end{thm}

\proof
Let $\seq{\phi_e}_{e \in \w}$ be an effective listing of all partial
recursive functions, and let $\seq{\Psi_e}_{e\in\w}$ be an effective
listing of all Turing functionals.  We let~$\psi_e$ denote the use
functional of~$\Psi_e$.  We make the assumption that for any oracle
$X$, if $\Psi_e^X(y)$ converges at stage $s$, then both $y$ and
$\psi_e^X(y)$ are less than $s$.

We build an r.e.\ set~$A$, an $A$-recursive function~$\Gamma^A$, and
partial recursive functions $\seq{f_e^0}_{e\in\w}$ and
$\seq{f_e^1}_{e\in\w}$ meeting the following requirements:
\begin{description}
\item[$G$] $\Gamma^A$ is total.
\item[$P_e$] If~$\phi_e$ is total, $\Gamma^A$ dominates~$\phi_e$.
\item[$R_e$] $\Psi_e^A$ fails to pdominate at least one of~$f_e^0$ and~$f_e^1$.
\end{description}
$G$ with the $P_e$'s guarantees that $A$ is high, while the $R_e$'s
ensure that $A$ is not pdominant.  The $R_e$-requirement will not be
met directly, but will instead be met through subrequirements:
\begin{description}
\item[$R_{e,n}$] There is an $i < 2$ and an~$x > n$ with $x \in
\hbox{dom} f_e^i$ such that for every~$y \leq x$ with $y \in
\hbox{dom} \Psi_e^A$, $\Psi_e^A(y) < f_e^i(x)$.
\end{description}
Clearly if infinitely many of these subrequirements are met, the
$R_e$-requirement will be met.

Our strategy for meeting an $R_{e,n}$-requirement will involve
guessing which of the~$\phi_{e'}$ are total, which we accomplish via a
tree of strategies.  We define the tree by devoting all nodes at
level~$2e$ to the $P_e$-requirement, and all nodes at
level~$2\seq{e,n}+1$ to the $R_{e,n}$-requirement.  The global
$G$-requirement does not appear on the tree.

Nodes devoted to a $P_e$-requirement have two possible
outcomes,~$\inff$ and~$\fin$ (denoting whether or not~$\phi_e$ is
total).  All other nodes have only one possible outcome, which we will
simply call $\outcome$.  Thus we can define the tree of strategies
recursively: the empty string $\seq{}$ is the root of the tree; if
$\sigma$ is on the tree at level~$2e$, then $\sigma\inff$ and
$\sigma\fin$ are its two children; if $\sigma$ is on the tree at level
$2\seq{e,n}+1$, then $\sigma\outcome$ is its unique child.

If $\tau_0$ and $\tau_1$ are nodes on the tree, we say that $\tau_0$
is to the left of $\tau_1$ if there is a node $\sigma$ on the tree
with $\sigma\inff \subseteq \tau_0$ and $\sigma\fin \subseteq \tau_1$.
 It is simple to verify that this is a transitive relation.

For a node~$\sigma$, accessible at stage~$s$ and of length $\leq s$,
we describe what actions~$\sigma$ takes at stage~$s$, and if $|\sigma|
< s$, which outcome of~$\sigma$ is next accessible.  We also describe
the action of the $G$-strategy.

\medskip
\noindent
{\bf Strategy for meeting the $G$-requirement.}
We partition~$\w$ in some effective fashion as $\w = \{ \seq{e,x} \mid
e < x\}$ with $\seq{e,x} < \seq{e,x'}$ whenever $x < x'$.  At
stage~$s$, for every $x < s$, if~$\Gamma^A(x)\, [s]$ is not defined, we define
\[
\Gamma^A(x)\, [s] = \max\{ \phi_{e,s}(x) \mid e < x \ \wedge \
\phi_{e,s}(x)\converge\}
\]
with use $\gamma^A(x)\, [s] = \max\{ \seq{e,x} \mid e < x\} + 1$.

\medskip
\noindent
{\bf Strategy for meeting the $P_e$-requirement.}
Whenever this strategy is initialised, we choose a threshold~$m$.  It
suffices to let $m$ be the stage at which the strategy was
initialised.  Let $\ell_s(e)$ be largest such
that~$\phi_{e,s}\uhr{\ell_s(e)}$ converges.

At stage~$s$, consider all $x < \ell_s(e)$.  If $\seq{e,x} > m$, we
enumerate~$\seq{e,x}$ into~$A$.  After considering all such~$x$, we
pass to the next accessible node.

Let $t < s$ be the last stage at which~$\sigma\inff$ was accessible
($t = 0$ if there is no such stage).  If $\ell_t(e) < \ell_s(e)$, we
let~$\sigma\inff$ be accessible and initialise all strategies $\tau
\supseteq \sigma\fin$.  Otherwise, we let $\sigma\fin$ be accessible.

\medskip
\noindent
{\bf Strategy for meeting the $R_{e,n}$-requirement.}
When~$\sigma$ is initialised, it will claim an interval of the form
$[k, 3k]$ on which it will define~$f_e^0$ and~$f_e^1$.  We must have
$k > n$, and the interval must be disjoint from any interval
previously claimed by any $R_{e',n'}$-requirement.  We can take this
as our definition of $k$: $k$ is the least number greater than $n$
with $[k,3k]$ disjoint from every interval previously claimed by any
$R_{e',n'}$-requirement.  Since only finitely many intervals can have
been claimed by any stage, there will always be such a $k$.
At a stage~$s$ when~$\sigma$ is accessible, let
\begin{eqnarray*}
b &=& \min\{\seq{e',\ell_s(e')} \mid \tau\inff \subseteq \sigma \hbox{
for some $P_{e'}$-strategy $\tau$}\};\\
p &=& \max\{\Psi_e^A(y)\, [s] \mid y \in \dom \Psi_e^A\, [s] \ \wedge
\ y \leq 3k \ \wedge \ \psi_e^A(y)\, [s] < b\};\\
q &=& \max\{f^i_{e,s}(x) \mid i < 2 \ \wedge \ k \leq x \leq 3k \
\wedge \ x \in \dom f_{e,s}^i\}.
\end{eqnarray*}
If $p < q$, we pass to the next accessible node.

Otherwise, let $i < 2$ and $x \in [k, 3k]$ be such that
$f_{e,s}^i(x)\diverge$ (we will later argue that such~$i$ and~$x$
exist).  We define $f_{e,s}^i(x) = p+1$, initialise all $\tau \supset
\sigma$ and pass to the next accessible node.

Since~$\sigma$ has only one possible outcome, this unique child is the
next accessible node.

\medskip
\noindent
{\bf Construction.}
We begin stage~$s$ by letting~$\seq{}$ be accessible and acting
according to its described strategy.  This strategy determines one of
its children to be the next accessible node, which then acts according
to its strategy.  We continue in this fashion, letting each accessible
node choose one of its children to be the next accessible node until
we reach a node of length $s$.  Once we have acted according to the
strategy for the node of length $s$, we let the $G$-strategy act and
then proceed to stage $s+1$.

\medskip
\noindent
{\bf Verification.}
We define the true path of the construction recursively.  The empty
string $\seq{}$ is on the true path.  If $\sigma$ is on the true path
and $\sigma$ is devoted to some $R_{e,n}$-requirement, then
$\sigma\outcome$ is on the true path.  If $\sigma$ is on the true path
and $\sigma$ is devoted to some $P_e$-requirement, and $\sigma\inf$ is
accessible at infinitely many stages, then $\sigma\inf$ is on the true
path; otherwise, $\sigma\fin$ is on the true path.  Note that an
equivalent definition would be that the true path is formed by
choosing from every level of the tree the leftmost node which is
accessible at infinitely many stages.

We perform the verification as a sequence of claims.

\begin{clm}
If $\tau_0$ and $\tau_1$ are nodes on the tree of strategies with
$\tau_0$ to the left of $\tau_1$, and $\tau_0$ is accessible at stage
$s$, then $\tau_1$ is initialised at stage $s$
\end{clm}

\proof
Fix $\sigma$ with $\sigma\inff \subseteq \tau_0$ and $\sigma\fin
\subseteq \tau_1$.  Then $\sigma$ is a $P_e$-strategy that was
accessible at stage $s$ and chose $\sigma\inff$ to be the next
accessible node.  By construction, all nodes $\tau \supseteq
\sigma\fin$ are initialised at this stage, including $\tau_1$.\qed

\begin{clm}
$\Gamma^A$ is total.
\end{clm}

\proof
At every stage~$s$, we define $\Gamma^A(x)\, [s]$ for all $x < s$. 
Since~$A$ is r.e., and the use~$\gamma^A(x)\, [s]$ never increases,
$\Gamma^A(x)$ is defined for all~$x$.\qed

\begin{clm}\label{claim:growing domain}
If $\sigma$ is an $R_{e,n}$-requirement,~$\sigma$ has outcome $\inff$
at stage~$s$,~$\sigma$ is not initialised between stages~$s<t$, and $y
\leq 3k$ is such that $\Psi_e^A(y)\, [s]$ was included in the maximum
from which~$p$ was defined at stage~$s$, then $y \in \hbox{dom }
\Psi_e^A\, [t]$ and $\Psi_e^A(y)\, [t] = \Psi_e^A(y)\, [s]$.
\end{clm}

\proof
The only way the conclusion might fail is if some $P_{e'}$-strategy
were to enumerate an element into~$A$ below~$\psi_e^A(y)\, [s]$
between stages~$s$ and~$t$.  We consider the possibilities.

Since~$\sigma$ was not initialised, no $P_{e'}$-strategy to the left
of~$\sigma$ was accessible to do this.  Similarly, no
$P_{e'}$-strategy~$\tau$ with $\tau\fin \subseteq \sigma$ could
enumerate a new element into~$A$ without $\ell_s(e')$ increasing,
causing~$\tau\inff$ to be accessible and initialising~$\sigma$.

Any $P_{e'}$-strategy to the right of~$\sigma$ or extending~$\sigma$
would have been initialised at stage~$s$ and so its next threshold
would be no less than~$s$, which by our earlier assumption is greater
than~$\psi_e^A(y)\, [s]$.  So such a strategy cannot enumerate an
element below~$\psi_e^A(y)\, [s]$.

Any $P_{e'}$-strategy~$\tau$ with $\tau\inff \subseteq \sigma$ has
$\seq{e',\ell_s(e')} > \psi_e^A(y)\, [s]$, and thus has already
enumerated by stage~$s$ all the elements below~$\psi_e^A(y)\, [s]$
that it ever will.\qed

\begin{clm}
Let $\sigma$ be an $R_{e,n}$-requirement.  Suppose~$\sigma$ was
initialised at stage~$s_0$, choosing an interval~$[k,3k]$,
and~$\sigma$ was not initialised between stages $s_0 < t$.  Then there
is an $i < 2$ and an $x \in [k, 3k]$ such that $f_{e,t}^i(x)\diverge$.
\end{clm}

\proof
Since $[k, 3k]$ was chosen disjoint from every previously claimed
interval, and the functions~$f_e^i$ are only defined within a claimed
interval, if~$s$ is a stage with some $f_e^i(x)$ defined with $x \in
[k, 3k]$, then $s_0 \leq s$.  By Claim \ref{claim:growing domain}, at
every such stage $s \leq t$, the domain of $\Psi_e^A\, [s]$ grows by
at least one element which is no more than~$3k$ and which is still in
$\hbox{dom } \Psi_e^A\, [t]$.  Thus this can happen at most~$3k+1$
many times.  Since at most one $f_e^i(x)$ definition is made
by~$\sigma$ at each such stage $s$, and there are $4k+2$ many pairs
$(i, x)$ with $i < 2$ and $x \in [k, 3k]$, there is always a pair with
$f_{e,t}^i(x)\diverge$.\qed

\begin{clm}
If~$\sigma$ is a strategy along the true path, $\sigma$ is only
initialised finitely many times.
\end{clm}

\proof
By induction on~$|\sigma|$.  The case when $\sigma = \seq{}$ is trivial.

A strategy~$\sigma$ can only be initialised by the actions of
strategies $\tau \subset \sigma$.

If $\tau$ is a $P_e$-requirement, and $\tau\inff \subseteq \sigma$,
$\tau$ cannot initialise~$\sigma$.

If $\tau$ is a $P_e$-requirement, and $\tau\fin \subseteq \sigma$,
then $\tau\fin$ is on the true path.  By definition of the true path,
there is some stage after which $\tau$ never again has
outcome~$\inff$.  After this stage,~$\tau$ can never again initialise~$\sigma$.

If $\tau$ is an~$R_{e,n}$-requirement, let~$s_0$ be the last stage at
which~$\tau$ is ever initialised.  By Claim \ref{claim:growing
domain}, at every stage~$s > s_0$ at which $\tau$
initialises~$\sigma$, the domain of $\Psi_e^A\, [s]$ grows by at least
one element which is no more than~$3k$ and which never subsequently
leaves.  Thus this can happen at most~$3k+1$ many times.\qed

\begin{clm}
If $\phi_e$ is total, $\Gamma^A$ dominates~$\phi_e$.
\end{clm}

\proof
Let~$\sigma$ be the $P_e$-strategy along the true path.  Let~$s_0$ be
the final stage at which~$\sigma$ is initialised, and let~$m$ be the
threshold chosen at this stage.  For an~$x$ with $\seq{e,x} > m$,
let~$s$ be the least stage such that $\ell_s(e) > x$.  By
construction, $\seq{e,x}$ will not enter~$A$ before stage~$s$.

At the first stage $t \geq s$ when~$\sigma$ is accessible,~$\sigma$
will enumerate $\seq{e,x}$ into~$A$ if no other $P_e$-strategy has
already done so.  When this happens, the $G$-strategy will redefine
$\Gamma^A(x) \geq \phi_e(x)$.  By construction, all future
redefinitions of $\Gamma^A(x)$ will be at least~$\phi_e(x)$.  Thus
$\Gamma^A(x) \geq \phi_e(x)$ for all~$x$ with $\seq{e,x} > m$.\qed

\begin{clm}
For every~$n \in \w$, there is an~$i < 2$ and an~$x > n$ with $x \in
\dom f^i$ such that for every $y \leq x$ with $y \in \hbox{dom }
\Psi_e^A$, $\Psi_e^A(y) < f_e^i(x)$.
\end{clm}

\proof
Let $\sigma$ be the $R_{e,n}$-strategy along the true path.  Let $s_0$
be the final stage at which~$s$ is initialised, and let~$[k,3k]$ be
the interval chosen at stage~$s_0$.  Recall that we chose $k > n$. 
Let $i < 2$ and $x \in [k, 3k]$ be the final pair for which~$\sigma$
defines~$f_e^i(x)$.  We claim these are the desired values.

Suppose there were a $y \leq x$ with $y \in \hbox{dom } \Psi_e^A$ and
$\Psi_e^A(y) \geq f_e^i(x)$.  Let $s_1 > s_0$ be a stage such that $y
\in \hbox{dom } \Psi_{e,s_1}^A$ and $A_{s_1}\uhr{\phi_e^A(y)} =
A\uhr{\phi_e^A(y)}$.  Let $s_2 > s_1$ be a stage such that $\seq{e',
\ell_s(e')} > \phi_e^A(y)$ for every~$\tau\inf\subseteq \sigma$ a
$P_{e'}$-strategy.  Let $s_3 > s_2$ be a stage at which~$\sigma$ is accessible.

At stage~$s_3$, $p \geq \Psi_e^A(y) \geq f_e^i(x) = q$.  By
construction, a new pair~$i' < 2$ and $x' \in [k, 3k]$ will be chosen
and~$f_e^{i'}(x')$ defined, contradicting our choice of~$i$ and~$x$.\qed

\noindent
This completes the proof of Theorem \ref{T:5.3}.\qed

\section{The opposite of dominance}

\noindent
Hyperimmune-free degrees are in a certain way the counterpart of high
degrees: Every high set $A$ computes a function $f$ which grows faster
than every recursive function \cite{Ma66a,Ma66b};
every hyperimmune-free set $A$ only
computes functions $f$ for which there is a recursive function growing
faster than them (this recursive function depends on $f$).
One can say that besides being low also having hyperimmune-free Turing
degree is something like the opposite of being high. Then, every total
$A$-recursive function is majorised by a recursive one. One could ask
whether the same could be possible for the type of sets investigated
in the present work.

\begin{defi}
Call a set $A$ to be phif (hyperimmune-free with respect to partial
functions) iff for every partial $A$-recursive function $\psi$
there is a partial recursive function $\phi$ and a constant $k$
such that for every $x \in dom(\psi)$ there is a $y \in dom(\phi)$
with $y \leq kx$ and $\phi(y) > \psi(x)$.
\end{defi}

\begin{thm}
Every phif set $A$ has hyperimmune-free Turing degree.
\end{thm}

\proof
It is shown that a set $A$ of hyperimmune Turing degree cannot
be phif.

To show this, the following fact is used: There is an
$A$-recursive function $F$ such that for every infinite r.e.\
set $W_e$ and every $n$ there are infinitely many $m \in W_e$ with
$F(m) = n$.

One can define such a function $F$ inductively by
starting a monotonically increasing function $G \leq_T A$ not majorised
by any recursive function and then letting $F(m) = n$ for the least
triple $\langle e,n,k \rangle$ such that either $\langle e,n,k \rangle = m$
or $m \in W_{e,G(m)}$ and there are less than $k$ numbers $o < m$ with
$F(o) = n \wedge o \in W_{e,G(m)}$. It is clear that $F$ is a total
$A$-recursive function. Furthermore, a requirement given by a
triple $\langle e,n,k \rangle$ is satisfied when there are at least
$k$ numbers $o \in W_e$ with $F(o) = n$. One can see that whenever
$W_e$ is infinite then each requirement $\langle e,n,k \rangle$ becomes
eventually satisfied as there are infinitely many $m$ with
$m \in W_{e,G(m)}$ where $\langle e,n,k \rangle$ requires
attention unless the requirement is satisfied; here the $m$-th
requirement can only be denied attention when a higher order
requirement, that is the $0$-th, $1$-st, $\ldots$ or $m-1$-th
requirement is done and will be satisfied by defining $F$ at one
place accordingly. This happens of course only finitely often
and then the $m$-th requirement will be satisfied eventually.
So $F$ has the requested properties.

Given $F$, define a partial $A$-recursive function $\psi$ as follows.
On input $x$ with $2^m \leq x < 2^{m+1}$ and $F(m) = \langle i,j,k\rangle$,
$\psi(x)$ simulates the computations of $\varphi_i(y)$ for
inputs $y \leq 2^{m+1} \cdot k$ until at least $h = j \cdot 2^m +(x-2^m)$ of
these computations have converged with outputs $z_0,z_1,\ldots,z_{h-1}$.
Then $\psi^A(x)$ takes the value $1+z_0+z_1+\ldots+z_{h-1}$.

Now assume by way of contradiction that there is a partial recursive
function $\varphi_i$ and a $k$ such that for almost all $x \in dom(\psi)$
there is a $y \leq kx$ in the domain of $\varphi_i$ with
$\varphi_i(y) > \psi(x)$. Let $j$ be the lim sup for $m \rightarrow \infty$
of the numbers $j_m$ where each $j_m$ is the rounded-down value of
$2^{-m} \cdot |dom(\varphi_i)\cap \{0,1,\ldots,2^{m+1}\cdot k\}|$.
As $j_m \in \{0,1,2,\ldots,2k+1\}$ for each $m$, the lim sup of the
$j_m$ must exist and is a number $j$. The set $E = \{m: j_m = j\}$
is r.e.\ and infinite. Hence there are infinitely many $m \in E$
with $F(m) = \langle i,j,k \rangle$. For each of these $m$ there
is an $x$ such that $2^m \leq x < 2^{m+1}$ and
$j \cdot 2^m+(x-2^m) = |dom(\varphi_i) \cap \{0,1,\ldots,2^{m+1} \cdot k\}|$.
Now $\psi(x)$ is defined and $\psi(x) > \varphi_i(y)$ for every
$y \in dom(\varphi_i)$ with $y \leq 2^{m+1} \cdot k$. This produces the
desired contradiction and therefore $i,k$ cannot destroy the witness $\psi$
against $A$ being phif. Hence, every phif set is hif, that is,
has hyperimmune-free Turing degree.\qed

\medskip
\noindent
The next result shows that phif sets are even
recursive. Kummer's Cardinality Theorem \cite{HKO92,Ku92} implies the
following result, which confirmed a conjecture by Owings \cite{Ow89}.

\begin{thm}\label{lem_function_avoiding}
If a set $A$ of natural numbers, a natural number $k$ 
and a function $g$ with $k$ inputs satisfy
\begin{quote}
   $\forall x_1,x_2,\ldots,x_k \ [x_1 < x_2 < \ldots < x_k$
   $\Rightarrow$ \\ \mbox{ } \hfill{}
   $g(x_1,\ldots,x_k) \in \{0,1,\ldots,k\}-\{|A \cap \{x_1,\ldots,x_k\}|\}]$
\end{quote}
then $A \leq_T g$.
\end{thm}

\noindent
The next result uses this theorem now to prove that $A$ is recursive;
the proof first shows that $A$ is $K$-recursive and then invokes the
above result that $A$ has hyper\-im\-mune-free Turing degree in order to
show that $A$ is recursive.

\begin{thm}
If $A$ is phif, then $A$ must be recursive.
\end{thm}

\proof
Suppose that $A$ is phif. Let $\psi^A( 2^e + 2^{e+1}x) =
\phi_e^A(x)$. Let $p$ and $\ell$ be such that for every $x \in
\dom(\psi)$, there is a $y \in \dom(p)$ with $y \le \ell x$ and $p(y)
> \psi(x)$.

We will construct a partial $A$-recursive function $h = \phi_e^A$,
and by the usual recursion theorem argument we may assume we already
know $e$.  Let $k = 2^{e+2}\ell$ and let $f$ be an $A$-recursive function
with $f(m) = |A \cap \{x_1,\ldots,x_k\}|$ for the $m$-th tuple
$(x_1,\ldots,x_k)$ with respect to a recursive enumeration of all
$k$-tuples. For all sufficiently large $x
\in \dom(h)$, there is a $y \in \dom(p)$ with $y \le kx$ and $p(y) > h(x)$.

Construct a pair of sequences $\seq{b_n}_{n \in \w}$ and $\seq{c_n}_{n
\in \w}$ as follows:
\begin{itemize}
\item $b_0 = 1$;
\item $b_{n+1} = (k+1)b_n + 1$;
\item $c_0 = k$;
\item $c_{n+1} = (k+1)(b_{n+1} - b_n) - 1$.
\end{itemize}
The important properties of these sequences are the following:
$$
c_{n+1} < (b_{n+1} - b_n)(k+1) \mbox{ and }
c_{n+1} = kb_{n+1}.
$$
Let
$$
   g(n) = |\{m: m \in \dom(p) \wedge m < c_{n+1}\}| \% (k+1).
$$
That is, $g(n)$ is the residue when the number of elements in the
domain of $p$ less than $c_{n+1}$ is reduced mod $(k+1)$. Note that
$g$ is $K$-recursive.

We construct $h$ as follows: for $(b_n + i) \in [b_n, b_{n+1})$, wait
until $(k+1)i + f(n)$ many elements less than $c_{n+1}$ have entered
the domain of $p$.  Then define $h(b_n+i)$ greater than the largest
value of $p$ seen so far.

For any $n$ with $f(n) = g(n)$, let $i$ be such that
$$
   (k+1)i + f(n) = |\{m: m \in \dom(p) \wedge m < c_{n+1}\}|.
$$
Note that $i \le \frac{c_{n+1}}{k+1} < b_{n+1} - b_n$,
so $b_n \leq b_n + i < b_{n+1}$.  Then by construction, $h(b_n+i)$
is larger than $p(y)$ for all $y < c_{n+1}$.  So $h(b_n+i)$ is larger
than $p(y)$ for all $y < (b_n+i)k$.  By our choice of $p$ and $k$,
this can only happen finitely many times.  So $f(n) = g(n)$ for at
most finitely many $n$. By Theorem~\ref{lem_function_avoiding}, $A
\leq_T g$ and so $A \leq_T K$. As $A$ has also hyperimmune-free
Turing degree, it follows that $A$ is recursive.\qed

\begin{rem}
One might also look at the other opposite of highness, that is, the
analogue of lowness. The corresponding question would be: When is $K$
pdominant relative to $A$? More precisely, for which $A$ does it hold
that there is a partial $K$-recursive
function $\psi$ such that for all partial $A$-recursive functions $\phi$
and almost all $x \in dom(\phi)$ there is a $y \in dom(\psi)$ with
$y \leq x \wedge \psi(y) > \phi(x)$. Indeed, there are uncountably
many such $A$ as one could take all sets which are low for $\Omega$,
that is, all sets $A$ relative to which $\Omega$ is Martin-L\"of random.
Such sets then satisfy that there is no partial $A$-recursive function
$\phi$ satisfying that $\phi(x) > c_{\Omega}(x)$ for infinitely many
$x \in dom(\phi)$ where $c_{\Omega}$ is the convergence module of $\Omega$;
otherwise $\Omega$ would not be Martin-L\"of random relative to $A$.
Hence the $K$-recursive function $\psi(x) = c_{\Omega}(2x)$ witnesses
the corresponding lowness property.
\end{rem}

\section{Weak truth-table reducibility}

\noindent
Recall that a partial $A$-recursive function $\psi$ is wtt-reducible
to $A$ iff there is an index $e$ and a recursive function
$f$ such that $\psi(x) = \varphi^{\{y \in A: y < f(x)\}}_e(x)$ for all~$x$.
Now a function is {\em wttpdominant} iff there is a partial function
$\psi \leq_{wtt} A$ such that for every partial recursive $\phi$
and almost all $x \in dom(\phi)$ there is an
$y \in dom(\psi)$ with $y \leq x \wedge \psi(y) > \phi(x)$.
Arslanov showed that an r.e.\ set is wtt-complete iff it wtt-computes
a fixed-point free function. The next result shows that wttpdominance
is a further criterion for wtt-completeness of r.e.\ sets, which stands
in contrast to the situation at Turing reducibility.

\begin{thm}
An r.e.\ set is wttpdominant iff it is wtt-complete.
\end{thm}

\proof
Only the direction ``wttpdominant $\Rightarrow$ wtt-complete'' needs
to be shown. Let $\psi \leq_{wtt} A$ be given and let $f$ be the use
of the wtt-reduction. Now define that $\phi(x)$ with $2^m \leq x < 2^{m+1}$
takes the value $\psi^{A_s}(x-2^m)$ iff $s$ is the time which $m$ needs
to be enumerated into $K$; $\phi(x)$ is undefined when $m \notin K$.
By Proposition~\ref{pr:factork} there is for almost all $m$ and all
$x \in dom(\phi)$ with $2^m \leq x < 2^{m+1}$ a $y$ in the domain
of $\psi$ such that $y < 2^m$ and $\psi(y) > \phi(x)$. Given now any
$m$, let $y < 2^m$ be chosen such that $\psi(y)$ is the largest value
among these $y$ and $x = 2^m+y$. It follows from above property that
either $\phi(x)$ is undefined and $m \notin K$ or $\phi(x) < \psi(y)$
and therefore $A_s$ below $f(2^m)$ different from $A$ below $f(2^m)$.
Now define the $A$-recursive function $g$ with $g(m)$ being the first
time $s$ where $A_s$ equals $A$ below $f(2^m)$, one has for almost all
$m$ that $m \in K_{g(m)}$ iff $m \in K$. This shows that $A$ is
wtt-complete.\qed

\medskip
\noindent
Furthermore, one can show that no hyperimmune set and thus also no
$1$-generic set is wttpdominant. On the other hand, Example~\ref{ex:omega}
actually shows that the half $A$ of $\Omega$ is wttpdominant. Hence there
is a wtt-incomplete wttpdominant set $A \leq_{wtt} K$. Note that
the negative results on pdominance transfer to wttpdominance. 

\section*{Acknowledgments}

\noindent
The authors want to thank Noam Greenberg
for discussions and correspondence on this subject; furthermore
they are greatful to the anonymous referees for detailed and comprehensive
comments.

\end{document}